\documentclass{andp2012}
\usepackage[english]{babel}
\usepackage{epstopdf}

\usepackage{setspace}

\DeclareMathOperator{\sgn}{sgn}
\def\p{{\bf p}}

\def\r{{\bf r}}
\def\Dr{{\Delta\r}}

\def\a{{\bf a}}
\def\A{{\cal A}}
\def\one{{\bf 1}}

\newcommand{\bra}[1]{\langle #1 |}
\newcommand{\ket}[1]{| #1 \rangle}

\title{Coulomb electron pairing in a tight-binding model of La-based cuprate superconductors}
\author[K.\,M. Frahm]{K.\,M. Frahm,\inst{1}}
\author[D.\,L. Shepelyansky]{D.\,L. Shepelyansky\inst{1,}
\footnote{Corresponding author\quad E-mail:~\textsf{dima(at)irsamc.ups-tlse.fr}}}
\address[1]{Laboratoire de Physique Th\'eorique, IRSAMC, 
  Universit\'e de Toulouse, CNRS, UPS, 31062 Toulouse, France}
\begin{abstract}
We study the properties of two electrons with Coulomb interactions in 
a tight-binding model of La-based cuprate superconductors. This tight-binding model
is characterized by  long-range hopping obtained previously by advanced 
quantum chemistry computations. We show analytically and numerically that 
the Coulomb repulsion leads to a formation of compact pairs propagating 
through the whole system. The mechanism of pair formation is related
to the emergence of an effective narrow energy band for Coulomb 
electron pairs with conserved total pair energy and momentum. 
The dependence of the pair formation probability on
an effective filling factor is obtained with a maximum around 
a filling factor of 20 (or 80) percent. 
The comparison with the case of the nearest neighbor tight-binding model
shows that the long-range hopping provides an increase of the phase space 
volume with high pair formation probability. We conjecture that the 
Coulomb electron pairs discussed here may play a role in high temperature 
superconductivity.
\end{abstract}
\shortabstract
\begin{document}
\maketitle

\section{Introduction}
\label{sec1}
The phenomenon of high temperature superconductivity (HTC), discovered in \cite{hts1986},
still requires its detailed  physical understanding as discussed by various experts of this field
(see e.g. \cite{dagotto,kivelson,proust}). 
The analysis is complicated by the complexity of the phase diagram and 
strong interactions between electrons (or holes). As a generic model, that can be used for 
a description of most superconducting cuprates, it was proposed to use a simplified 
one-body Hamiltonian with nearest-neighbor hopping on 
a square lattice formed by the Cu ions \cite{anderson}.
In addition the interactions between electrons are considered as 
a strongly screened Coulomb interaction 
that results in the 2D Hubbard model \cite{anderson}.  
However, a variety of experimental results
cannot be described by the 2D Hubbard model (see e.g. discussion in \cite{fresard}). Other models of type Emery 
\cite{emery1,emery2,varma,loktev} were developed and extended on 
the basis of extensive computations with various numerical methods of quantum chemistry
(see e.g. \cite{markiewicz,fresard} and Refs. therein). These studies demonstrated the importance
of next-nearest hopping and allowed to determine reliably the
longer-ranged tight-binding parameters. 

In this work we use the 2D longer-ranged tight-binding parameters reported 
in  \cite{fresard} and study the effects of Coulomb interactions between 
electrons in the frame work of this tight-binding model. There are different 
reasons indicating that long-range interactions between electrons 
may lead to certain new features as compared to the Hubbard case 
(see \cite{kivelson,proust,fresard}). Recently, we demonstrated that for 
two electrons on a 2D lattice with nearest-neighbor hopping the energy and 
momentum conservation laws leads to appearance of an effective narrow 
energy band for energy dispersion of two electrons \cite{prr2020}. 
In such a narrow band even a repulsive Coulomb interaction leads to 
electron pairing and ballistic propagation of such pairs through the whole system.
The internal classical dynamics of electrons inside such a pair is chaotic 
suggesting nontrivial properties of pair formation in the quantum case.
In this work we extend the investigations of the properties of such Coulomb electron pairs
for a more generic longer-ranged tight-binding lattice of one-body Hamiltonian 
typical for La-based cuprate superconductors. We find that the long-range hopping leads to 
new features of Coulomb electron pairs.

In Sec.~\ref{sec2} a detailed description of the tight-binding 
model for two interacting electrons for general lattices with a 
particular application to HTC is presented together with an analysis 
of the effective band width at fixed conserved total pair momentum. 
Section~\ref{sec3} provides first results of the full space time evolution 
obtained in the frame work 
of the Trotter formula approximation. Section~\ref{sec4} 
introduces the theoretical basis for the description in terms of 
an effective block Hamiltonian for a given sector of fixed 
momentum of a pair with 
technical details provided in Appendix \ref{appA}. In Sec.~\ref{sec5} 
the phase diagram of the long time average of the 
pair formation probability in the plane of 
total momentum is discussed while Sec.~\ref{sec6} provides 
some results for the intermediate time evolution of pair formation. 
An overview of the results for the pair formation probability 
at different filling factors is given in 
Section~\ref{sec7}. The final discussion is presented in Section~\ref{sec8}.

\section{Generalized tight-binding model on a 2D lattice}
\label{sec2}

We assume that each electron moves on a square lattice 
of size $N\times N$ with periodic boundary conditions with respect to 
the following generalized one-particle tight-binding Hamiltonian:
\begin{equation}
\label{eq_H1p}
H_{1p}=-\sum_{\r}\sum_{\a\in\A} t_\a\,\bigl(\ket{\r}\bra{\r+\a}+
\ket{\r+\a}\bra{\r}\bigr)
\end{equation}
where the first sum is over all discrete lattice points $\r$ (measured 
in units of the lattice constant) and $\a$ belongs to a certain set of 
{\em neighbor vectors} $\A$ such that 
for each lattice state $\ket{\r}$ there are non-vanishing hopping matrix 
elements $t_\a$ with $\ket{\r+\a}$ and $\ket{\r-\a}$ for $\a\in\A$. 
To be more precise, due to notational reasons, we choose the set $\A$ to 
contain all neighbor vectors $\a=(a_x,a_y)$ in one half plane with 
either $a_x>0$ or $a_y>0$ if $a_x=0$ such that 
$\A'=\A\cup(-\A)$ is the {\em full set} of all neighbor vectors. 
For each vector $\a$ 
of the full set $\A'$, we require that any other vector $\tilde\a$ which 
can be 
obtained from $\a$ by a reflection at either the $x$-axis, $y$-axis or 
the $x$-$y$ diagonal also belongs to the full set $\A'$ and has the same 
hopping amplitude $t_\a=t_{\tilde\a}$. 

For the usual nearest neighbor tight-binding model (NN-model), 
already considered in \cite{prr2020}, we have 
the set $\A_{\rm NN}=\{(1,0),(0,1)\}$ with $t_{(1,0)}=t_{(0,1)}=t=1$. 
The numerical results presented in this work correspond either to the NN-model 
(for illustration and comparison) or to a longer-ranged tight-binding lattice 
according to \cite{fresard} which we denote as the HTC-model. For this case 
the set of neighbor vectors is 
$\A_{\rm HTC}=\{(1,0),(0,1),(2,0),(0,2),$
$(1,\pm 2),(2,\pm 1),(1,\pm 1), 
(2,\pm 2)\}$ and the hopping am\-pli\-tudes are: 
$t=t_{(1,0)}=1$, $t'=t_{(1,1)}=-0.136$, $t{''}=t_{(2,0)}=0.068$, 
$t{'''}=t_{(2,1})=0.061$ and $t^{(4)}=t_{(2,2)}=-0.017$ 
corresponding to the values given in Table 2 of \cite{fresard} (all energies 
are measured in units of the hopping amplitude $t=t_{(1,0)}=t_{(0,1)}$ 
which is therefore set to unity here; see also Fig.~6a of \cite{fresard}
for the neighbor vectors of the different hopping amplitudes).
The hopping amplitudes for other vectors such as $(0,1)$, $(1,-1)$, $(2,1)$, 
$(1,-2)$ etc. are obtained from the above amplitudes by the 
appropriate symmetry transformations, e.g. $t_{(1,-1)}=t_{(1,1)}=t'=-0.136$ 
etc.

Even though that most of our numerical results presented in this work 
apply to the HTC-model (or the NN-model), we 
emphasize that certain theoretical considerations given below, especially 
for the effective block Hamiltonian in relative coordinates at given 
total momentum, are valid for arbitrary generalized tight binding 
models with more general sets $\A$ and also with a potential generalization 
to other dimensions. 

The eigenstates of $H_{1p}$ given in (\ref{eq_H1p}) are simple plane waves:
\begin{equation}
\label{eq_state1p}
\ket{\p}=\frac1N\sum_\r\,e^{i\p\cdot\r}
\end{equation}
with energy eigenvalues:
\begin{equation}
\label{eq_en1p}
E_{1p}(\p)=-2\sum_{\a\in\A} t_\a\cos(\p\cdot \a)
\end{equation}
and momenta $\p=(p_x,p_y)$ such that $p_x$ and $p_y$ are integer multiples 
of $2\pi/N$ (i.e. $p_\alpha=2\pi l_\alpha/N$, $l_\alpha=0,\ldots,N-1$, 
$\alpha=x,y$). For the HTC model, we can give a 
more explicit expression of the energy dispersion:
\begin{equation}
\begin{aligned}
E_{1p}&(p_x,p_y) = - 2\left[\cos(p_x)+\cos(p_y)\right] \\
&-4t' \cos(p_x) \cos(p_y) - 2t{''} \left[\cos(2p_x)+\cos(2p_y)\right] \\
&-4t{'''} \left[\cos(2p_x) \cos(p_y) + \cos(2p_y) \cos(p_x)\right] \\
&-4t^{(4)} \cos(2p_x) \cos(2p_y)
\end{aligned}
\label{2dlattice}
\end{equation}
which corresponds to eq. (30) of \cite{fresard} (assuming 
$t=1$ and $t^{(5)}=t^{(6)}=t^{(7)}=0$).

The quantum Hamiltonian of the model with two interacting 
particles (TIP) has the form:
\begin{equation}
\label{eq_quant_Ham}
H=H_{1p}^{(1)}\otimes \one^{(2)}+\one^{(1)}\otimes H_{1p}^{(2)}+
\sum_{\r_1,\r_2}\bar U(\r_2-\r_1)\ket{\r_1,\r_2}\bra{\r_1,\r_2}
\end{equation}
where $H_{1p}^{(j)}$ is the one-particle Hamiltonian (\ref{eq_H1p}) 
of particle $j=1,2$ with positional coordinate $\r_j=(x_j,y_j)$ and 
$\one^{(j)}$ is the unit operator of particle $j$. 
The last term in (\ref{eq_quant_Ham}) represents a (regularized) 
Coulomb type long-range interaction 
$\bar U(\r_2-\r_1)=U/[1+r(\r_2-\r_1)]$ 
with amplitude $U$ and the effective distance 
$r(\r_2-\r_1)=\sqrt{\Delta\bar x^2+\Delta\bar y^2}$ between the 
two electrons on the lattice with periodic boundary conditions. 
(Here $\Delta\bar x = \min(\Delta x,N-\Delta x)$;  
$\Delta\bar y = \min(\Delta y,N-\Delta y)$;
$\Delta x = x_2-x_1$; $\Delta y = y_2-y_1$ and the latter differences 
are taken modulo $N$, i.e. $\Delta x=N+x_2-x_1$ if $x_2-x_1<0$ 
and similarly for $\Delta y$). 
Furthermore, we consider symmetric (spatial) wavefunctions with respect to 
particle exchange assuming an antisymmetric spin-singlet state 
(similar results are obtained for antisymmetric wavefunctions).

In absence of interaction ($U=0$) the energy eigenvalues 
(the classical energy) of the 
two electron Hamiltonian (\ref{eq_quant_Ham}) (the two electrons) at 
given momenta $\p_1$ and $\p_2$ are (is) given by:
\begin{equation}
\begin{aligned}
\label{eq_Ec}
E_c(\p_1,\p_2)&=E_{1p}(\p_1)+E_{1p}(\p_2)\\
&=-4\sum_{\a\in\A} t_\a\cos(\p_+\cdot \a/2)\cos(\Delta\p\cdot \a)
\end{aligned}
\end{equation}
where $\p_+=\p_1+\p_2$ is the total momentum and 
$\Delta \p=(\p_2-\p_1)/2$ is the momentum associated to the relative 
coordinate $\Delta\r=\r_2-\r_1$. For the NN-model Eq. (\ref{eq_Ec}) becomes 
$E_c(\p_1,\p_2)=-4\sum_{\alpha=x,y} \cos(p_{+\alpha}/2)\cos(\Delta p_\alpha)$. 

Due to the translational invariance the total momentum $\p_+$ is conserved 
even in the presence of interaction ($U\neq 0$) and only two-particle plane 
wave states with identical $\p_+$ are coupled by non-vanishing interaction 
matrix elements. For the case of the NN-model, analyzed in \cite{prr2020},
the kinetic energy at fixed $\p_+$ is 
bounded by $\Delta E_b=4\sum_{\alpha=x,y} |\cos(p_{+\alpha}/2)|$.
Thus for TIP states with $E>\Delta E_b$ the two electrons 
cannot separate and  propagate as one pair even if their
interaction is repulsive.
For $p_{+x}=p_{+y}=\pi+\delta$  being 
close to $\pi$  and $|\delta|\ll 1$ there are compact Coulomb electron pairs
even for very small interactions $U $ as soon as 
$\Delta E_b \approx 4|\delta|< U \ll B_2$ with $B_2=16+U$ being the maximal 
energy bandwidth\footnote{In the following we use the notation 
$B_2=16+U$ for the bandwidth of the NN-model.} in 2D. 
Thus the conservation of the total momentum 
of a pair with $p_{+x}=p_{+y} \approx \pi$ 
leads to the appearance of an effective narrow energy band 
with formation of coupled electron pairs propagating 
through the whole system. However, the results obtained in \cite{prr2020}
show that even for other values of $p_{+x}, p_{+y} $
the probability of pair formation is rather high.

\begin{figure}
  \begin{center}
    \includegraphics[width=0.9\columnwidth]{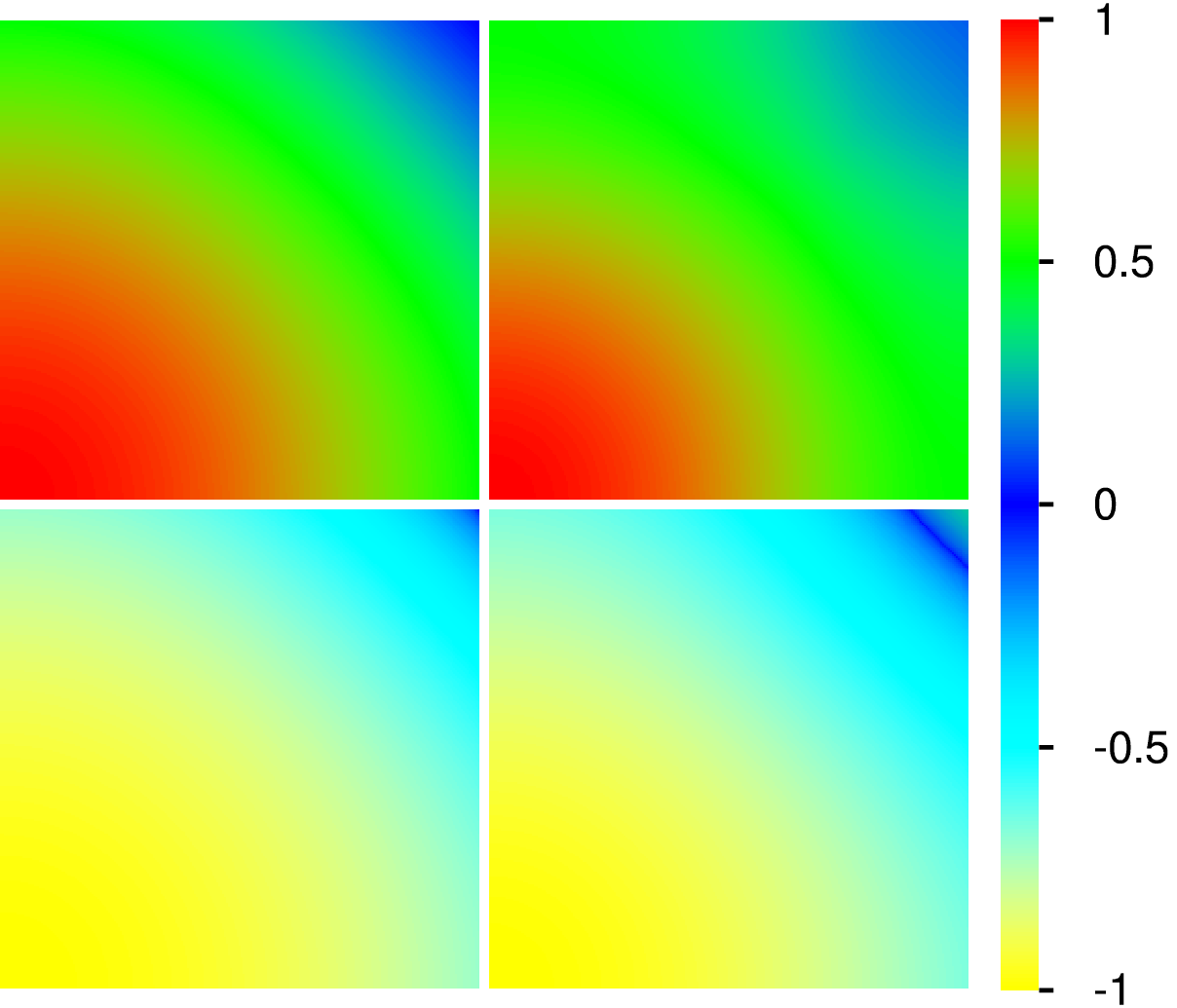}%
  \end{center}
  \caption{\label{fig1}
    Top panels show the 
    dependence of the effective electron pair band width 
    $\Delta E_b(\p_+)$ 
    on the pair momentum $\p_+=(p_{+x},p_{+y})$. Bottom panels show 
    the kinetic electron pair energy $E_c(\p_1,\p_2)$ 
    (in absence of interaction) 
    at momenta $\p_1=\p_2=\p_+/2$. Left panels correspond to 
    the NN-model and right panels to the HTC-model.
    In all panels the horizontal axis corresponds to $p_{+x}\in[0,\pi]$ 
    and the vertical axis to $p_{+y}\in[0,\pi]$.
    The numbers of the color bar correspond for top panels to the 
    ratio of the bandwidth over its maximal value and for lower panels to 
    the quantity $\sgn(E_c)\sqrt{|E_c|/E_{c,{\rm max}}}$ with 
    $E_{c,{\rm max}}$ being the maximum of $|E_c|$. 
    In all subsequent color plot figures the numerical values of the 
    color bar corresponds to the 
    ratio of the shown quantity over its maximal value. 
 }
\end{figure}

For the NN-model the effective band width for pairs $\Delta E_b $ can be
exactly zero for the specific pair momentum $\p_+=(\pi,\pi)$. 
However, this is not the case
for the HTC-model where due to the longer-ranged hopping
the minimal width $\Delta E_b$ is finite due to the additional terms 
with factors $\cos(\p_+\cdot\a/2)$ in (\ref{eq_Ec}). 
Therefore, we determined numerically for each given value of total 
momentum $\p_+$ the effective bandwidth as:
\begin{equation}
\label{eq_dEb}
\Delta E_b(\p_+)=\max_{\Delta\p}\left[E_c(\p_1,\p_2)\right]-
\min_{\Delta\p}\left[E_c(\p_1,\p_2)\right]
\end{equation}
with $\p_1=\p_+/2-\Delta \p$ and $\p_2=\p_+/2+\Delta \p$.
Top panels of Fig.~\ref{fig1} show density color plots of 
$\Delta E_b(\p_+)$ for the NN- and the HTC-model. 
For the HTC-case $\Delta E_b(\p_+)$ is maximal at $\p_+=(0,0)$ 
with value $\Delta E_{b,\rm max}=17.952$ and minimal 
at $\p_+=(\pi,\pi)$ with value $\Delta E_{b,\rm min}=2.176$ 
while for the NN-model we have $\Delta E_{b,\rm max}=16$ at $\p_+=(0,0)$ 
and $\Delta E_{b,\rm min}=0$ at $\p_+=(\pi,\pi)$. 
The value $\Delta E_{b,\rm min}=2.176$ for the HTC-model is still rather small
compared to the maximal value $\Delta E_{b,\rm max}\approx 18$ and we may 
expect a somewhat stronger pair formation probability for total momenta 
$\p_+$ close to $(\pi,\pi)$. However, this situation is qualitatively 
different as compared to the NN-model and the HTC-case 
requires new careful studies. 

For comparison, we also show in the lower panels of Fig.~\ref{fig1} 
the kinetic energy 
$E_c$ at $\p_1=\p_2=\p_+/2$ (for the square $\p_+\in[0,\pi]\times[0,\pi]$)
corresponding to $\Delta\p=0$. While for the 
NN-model this quantity vanishes at $\p_+=(\pi,\pi)$ there is for 
the HTC-model a zero-line between the two points $(\beta\pi,\pi)$ and 
$(\pi,\beta\pi)$ where $\beta\approx 0.877\approx 7/8$ is a numerical constant 
slightly below unity. 

\section{Full space time evolution of electron pairs}
\label{sec3}

As in \cite{prr2020} the full time evolution of two electrons is computed 
numerically for $N=128$ using the Trotter formula approximation  
(see  e.g. \cite{prr2020,trotter} for computational details).
We use the Trotter time step $\Delta t = B_2=1/(16+U)$ which is the 
inverse bandwidth for the case of NN-model. 
A further decrease of the time step does not affect the obtained results.
At the initial time both electrons are localized approximately 
at $ (N/2,N/2)$ with the distance 
$\Delta\bar x=\Delta\bar y=1$ using a linear combination of 8 states 
with all combinations due to particle exchange symmetry and 
reflection symmetry at the $\Delta x$- and $\Delta y$-axis. 
The method provides for each time value a wavefunction 
$\psi(x_1,y_1,x_2,y_2)$ from which we extract different quantities such 
as the density in $x_1$-$x_2$ plane:
\begin{equation}
\label{eq_xx2Ddens}
\rho_{XX}(x_1,x_2)=
\sum_{y_1,y_2}|\psi(x_1,y_1,x_2,y_2)|^2
\end{equation}
or the density $\Delta x$-$\Delta y$ plane:
\begin{equation}
\label{eq_dxdy2Ddens}
\rho_{\rm rel}(\Delta x,\Delta y)=
\sum_{x_1,x_2}|\psi(x_1,y_1,x_1+\Delta x,y_1+\Delta y)|^2
\end{equation}
(with position sums taken modulo $N$). We also compute the quantity 
$w_{10}$ by summing the latter density (\ref{eq_dxdy2Ddens}) over all 
values such that $|\Delta\bar x| \le 10$ and $|\Delta\bar y| \le 10$ which 
corresponds to a square of size $21\times 21$ in $\Delta x$-$\Delta y$ plane 
(due to negative values of $x_2-x_1$ etc.). 
This quantity gives the quantum probability to find both electrons at a 
distance $\le 10$ (in each direction) and we will refer to it as the 
{\em pair formation probability}. 

\begin{figure}
  \begin{center}
  \includegraphics[width=0.9\columnwidth]{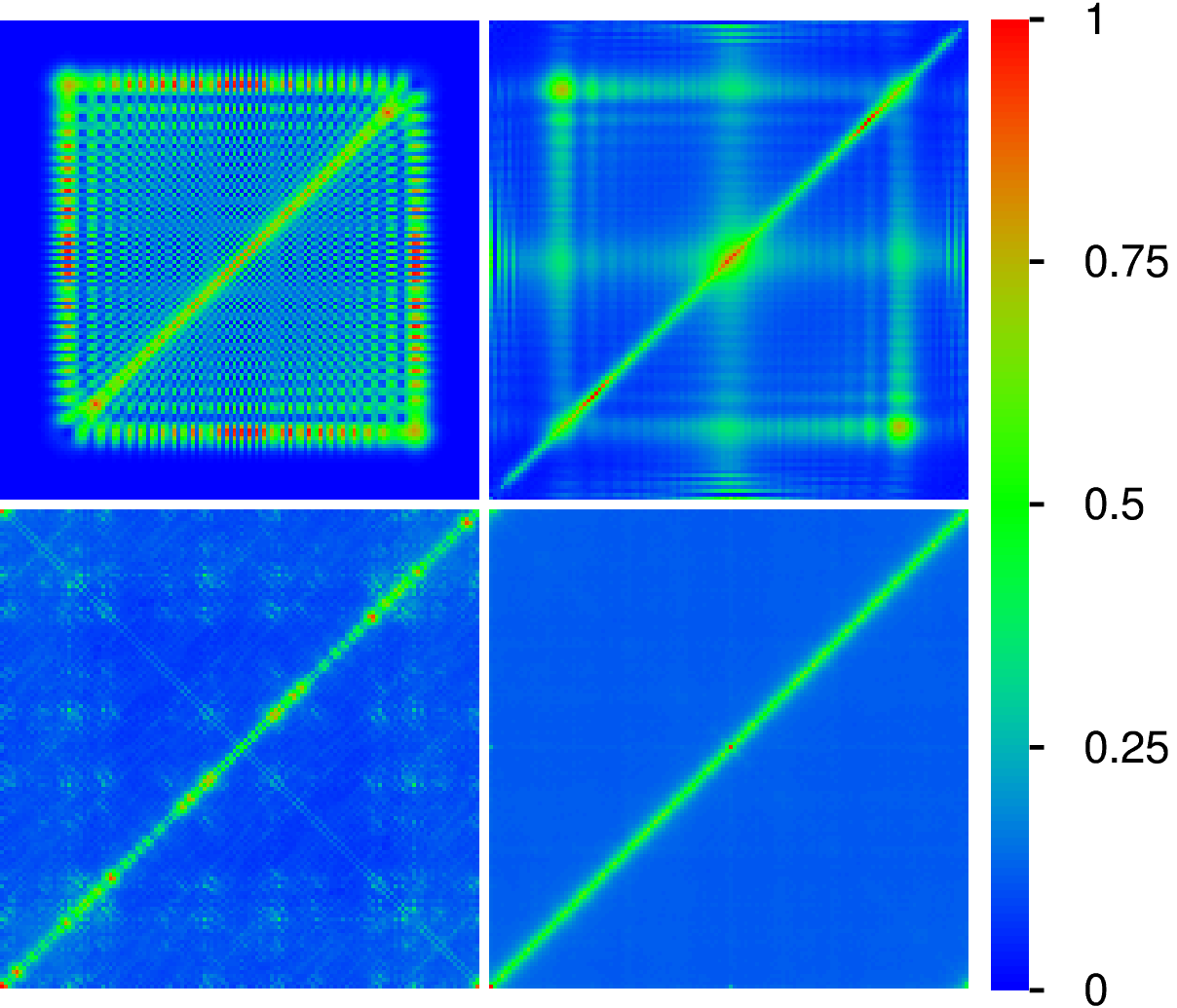}%
  \end{center}
  \caption{\label{fig2}
    2D Wavefunction density $\rho_{XX}(x_1,x_2)$ 
    in $x_1$-$x_2$ plane (see Eq. (\ref{eq_xx2Ddens})) 
    obtained from the time evolution 
    using the Trotter formula approximation for initial electron positions 
    at $\approx (N/2,N/2)$ with distance $\Delta\bar x=\Delta\bar y=1$ 
    for $N=128$, $U=2$ 
    and Trotter integration time step $\Delta t =1/B_2=1/(16+U)$. 
    Top (bottom) panels correspond to the time value 
    $t=445\,\Delta t$ ($t=10^4 \Delta t$) and 
    left (right) panels correspond to the NN-lattice (HTC-lattice). 
    The corresponding values of the pair formation probability 
    $w_{10}$ are 0.106 (top left), 0.133 (bottom left), 
    0.0940 (top right) and 0.125 (bottom right). 
    Related videos are available at \cite{suppmat,ourwebpage}.
 }
\end{figure}

In Fig.~\ref{fig2} the density $\rho_{XX}$ is shown for $U=2$,  
both NN- and HTC-models at two time values $t=445\Delta t$ and 
$t=10^4\,\Delta t$. 
These results show that the wavefunction has a component with electrons 
separating from each other and a component where electrons stay close to each 
other forming a pair propagating through the whole system 
that corresponds to a high density near a diagonal with $x_1 \approx x_2$. 
For $t=445\Delta t$ the value of $w_{10}$ is roughly 10\% and for 
$t=10^4\,\Delta t$ it is roughly 13\% for both models. However, the remaining 
diffusing component of about 87-90\% probability has a stronger periodic 
structure for the NN-model as compared to the HTC-model. 

\begin{figure}
  \begin{center}
  \includegraphics[width=0.9\columnwidth]{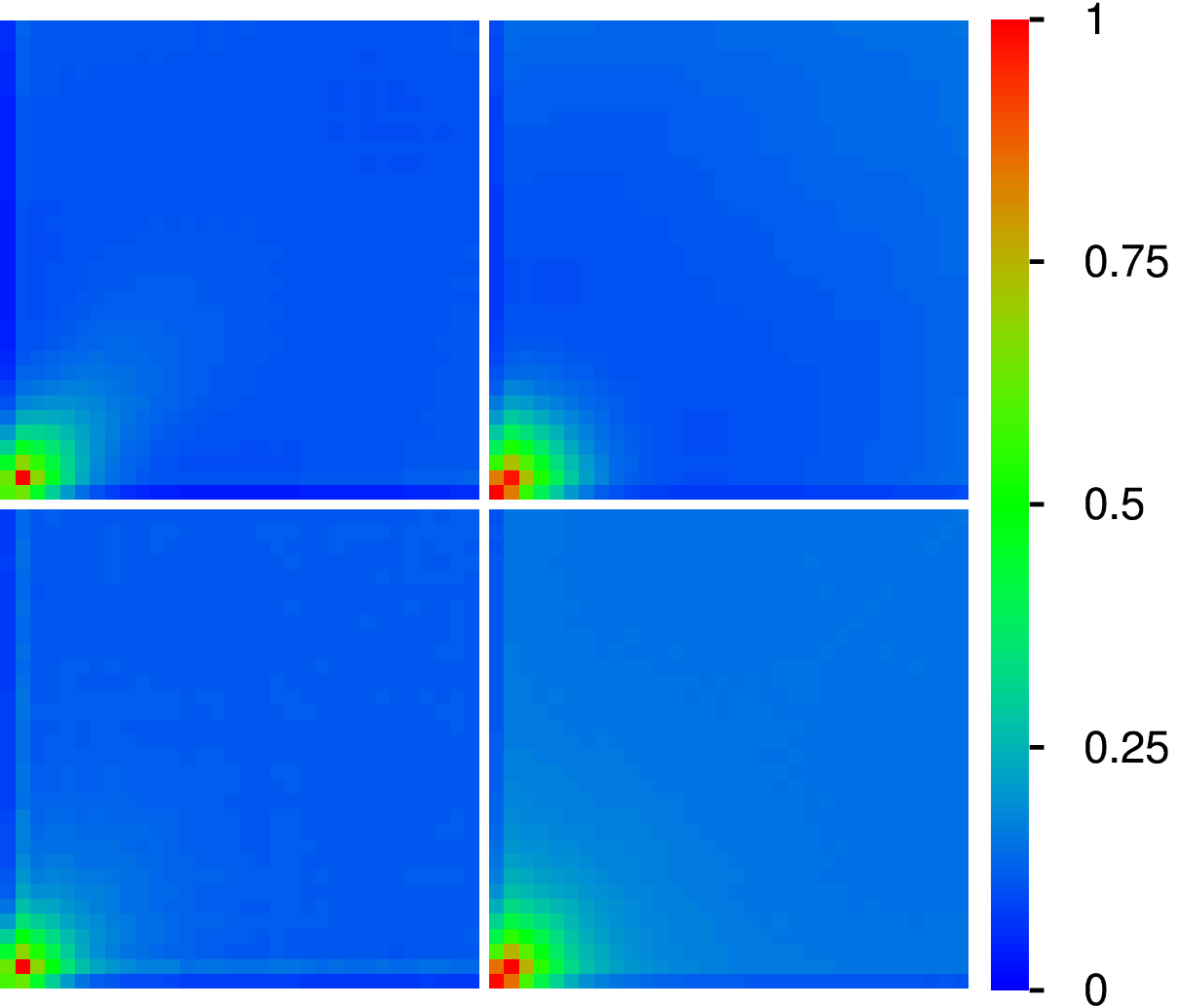}%
  \end{center}
  \caption{\label{fig3}
    2D Wavefunction density $\rho_{\rm rel}(\Delta x,\Delta y)$ 
    in $\Delta x$-$\Delta y$ plane of relative 
    coordinates (see Eq. (\ref{eq_dxdy2Ddens})) 
    for the same states, cases and parameters of Fig.~\ref{fig2}. 
    All panels show the zoomed density for $0\le \Delta x,\Delta y<32$. 
    Related videos are available at \cite{suppmat,ourwebpage}.
 }
\end{figure}

Figure~\ref{fig3} shows the density $\rho_{\rm rel}(\Delta x,\Delta y)$ 
for the same cases of Fig.~\ref{fig2}. We clearly see a strong enhancement 
of the probability at small values $\Delta\bar x\approx \Delta\bar y<5$ 
($<6-7$) for the NN-model (HTC-model) showing that there is a considerable 
probability that both electrons stay close to each other forming a 
Coulomb electron pair. Furthermore, the remaining 
wavefunction component of independently propagating electrons, clearly 
visible in Fig.~\ref{fig2}, is not visible in the density shown 
in Fig.~\ref{fig3} even though this component corresponds to 87-90\% 
probability. 

The supplementary material contains two videos (for $\sim 460$ 
time values in the range $\Delta t\le t\le 10^4\Delta t$ with 
roughly uniform logarithmic density) of the two densities 
$\rho_{XX}$ and $\rho_{\rm rel}$ where both models NN and HTC are directly 
compared in the same video. The raw-data used for these videos 
is the same as in Figs. \ref{fig2} and \ref{fig3}. 

\section{Time evolution in sectors of fixed total momentum}
\label{sec4}

As already mentioned in Sec.~\ref{sec3} the total momentum $\p_+$ is 
conserved by the TIP dynamics of the 
Hamiltonian (\ref{eq_quant_Ham}). In order to exploit 
this more explicitly, we introduce as in \cite{prr2020}, 
{\em block basis states} by:
\begin{equation}
\label{eq_block_basis}
\ket{\p_+,\Dr}=\frac1N\sum_{\r_1}e^{i\p_+\cdot(\r_1+\Dr/2)}
\ket{\r_1,\r_1+\Dr}
\end{equation}
where $\p_+=(p_{+x},p_{+y})$ (with $p_{+\alpha}=2\pi l_{+\alpha}/N$; 
$l_{+\alpha}=0,\ldots,N-1$; $\alpha=x,y$) is a fixed value of the total 
momentum and $\r_1$, $\Dr$ are vectors on the square lattice (with 
position sums in each spatial direction taken modulo $N$). One can show 
(see Appendix \ref{appA} for details) 
that the TIP Hamiltonian (\ref{eq_quant_Ham}) applied to such state gives a 
linear combination of such states for different $\Dr$ values but 
the {\bf same} total momentum value $\p_+$ which provides for each value or 
{\em sector} of $\p_+$ an effective {\em block Hamiltonian}:
\begin{equation}
\label{eq_block_ham}
\begin{aligned}
\bar h^{(\p_+)}=&
-\sum_{\Dr}\sum_{\a\in\A}
\bar t^{(\p_+)}_\a\bigl(\ket{\Dr+\a}\bra{\Dr}+
\ket{\Dr}\bra{\Dr+\a}\bigr)\\
&+\sum_{\Dr} \bar U(\Dr)\ket{\Dr}\bra{\Dr}
\end{aligned}
\end{equation}
where $\bar t^{(\p_+)}_\a=2\cos(\p_+\cdot\a/2)\,t_\a$ is an 
{\em effective rescaled hopping 
amplitude} depending also on $\p_+$ and we have for simplicity omitted the 
index $\p_+$ in the block basis states. This effective block Hamiltonian 
corresponds to a tight-binding model in 2D of similar structure 
as (\ref{eq_H1p}) with modified hopping amplitudes and 
an additional ``potential'' $\bar U(\Dr)$. We note that in absence of 
this external potential ($U=0$) the eigenfunctions of (\ref{eq_block_ham}) 
are plane waves and we immediately recover the expression (\ref{eq_Ec}) 
for its energy eigenvalues where $\Delta\p$ is the momentum associated to the 
relative coordinate $\Dr$. For the simple 
NN-model the result for the effective block Hamiltonian 
was already given in \cite{prr2020} and the above 
expression (\ref{eq_block_ham}) provides the generalization to arbitrary 
tight-binding lattices characterized by a certain set of neighbor vectors $\A$ 
and associated hopping amplitudes $t_\a$ (the generalization to arbitrary 
spatial dimension is also obvious). 
As already discussed in \cite{prr2020} the boundary conditions 
of (\ref{eq_block_ham}) 
in $x-$ ($y-$)direction are either periodic if the integer index $l_{+x}$ 
($l_{+y}$) of $p_{+x}$ ($p_{+y}$) is even or anti-periodic if this index is 
odd. This can be understood by the fact that the expression 
(\ref{eq_block_basis}) is modified by the factor 
$e^{\pm ip_{+x}N/2}=e^{\pm i\pi l_{+x}}=(-1)^{l_{+x}}$ if 
$\Delta x$ is replaced by 
$\Delta x\pm N$ and similarly for $\Delta y$ (with $\Dr=(\Delta x,\Delta y))$. 

Diagonalizing the effective block Hamiltonian (\ref{eq_block_ham}), we can 
rather efficiently compute the exact quantum time evolution 
$\ket{\bar\psi(t)}=e^{-i \bar h^{(\p+)}\,t}\,\ket{\bar\psi(0)}$
inside a given sector of $\p_+$. 
As initial state $\ket{\bar\psi(0)}$ 
we choose a state (in the reduced block space) 
given as the totally symmetric 
superposition of four localized states where $\Delta x$ and $\Delta y$ 
are either 1 or $N-1$. Such a state corresponds in full space to a plane wave 
in the center of mass direction with total fixed momentum $\p_+$ and strongly 
localized in the relative coordinate $\Dr$. The matrix size of 
(\ref{eq_block_ham}) is $N^2$ which corresponds to a complexity of $N^6$ for 
the numerical diagonalization. 

However, for a general lattice, such as the HTC-model, 
one can exploit the particle exchange symmetry 
to reduce the effective matrix size to roughly $N^2/2$
and for the special cases of $p_{+x}=p_{+y}$ or either 
$p_{+x}=0$ or $p_{+y}=0$ a second symmetry allows a further reduction 
of the effective matrix size to $\approx N^2/4$ 
(for the NN-model there are two or three symmetries for these cases 
with effective matrix sizes of $\approx N^2/4$ or $\approx N^2/8$ 
respectively; see \cite{prr2020} and Appendix \ref{appA} for details). 

In view of this, we have been able to compute numerically 
the exact time evolution for the HTC-model 
in certain $\p_+$ sectors for a lattice size up to $N=384$ for the case of two 
symmetries and a limited number of different other parameters 
(values of $\p_{+}$ and $U$). For the case of one symmetry 
and the exploration of all possible values of $p_{+x}$ and $p_{+y}$ we used 
the maximum system size $N=192$. We also implemented more expensive 
computations where no or less possible symmetries are used to verify (at 
smaller values of $N$) that they provide identical numerical results. 

We compute the wavefunction in block representation $\bar\psi(\p_+,\Dr)$ 
for about 700 time values $t=0$ and $10^{-1}\le t/\Delta t\le 10^6$ 
(with a uniform density in logarithmic scale) 
where $\Delta t=1/B_2=1/(16+U)$ is the time step already used for the Trotter 
formula approximation given as the inverse bandwidth for the case of 
the NN-model which is the smallest time (inverse of the largest energy) 
scale of the system. 

From the wavefunction we extract in a similar way as in Sec.~\ref{sec3} 
the pair formation probability $w_{10}$ by summing the (normalized) 
wavefunction density $|\bar\psi(\p_+,\Dr)|^2$ at fixed $\p_+$ 
over the $21\times 21$ square with $|\Delta\bar x| \le 10$ and 
$|\Delta\bar y| \le 10$. We also compute the inverse participation ratio:
\begin{equation}
\label{eq_ipr}
\xi_{\rm IPR}=\left(\sum_\Dr |\bar\psi(\p_+,\Dr)|^4\right)^{-1}
\end{equation}
which gives roughly the number of lattice sites (in $\Dr$ space) 
over which the wavefunction 
is localized. Both quantities $w_{10}$ and $\xi_{\rm IPR}$ converge typically 
rather well to their stationary values 
at times $t>10^3\Delta t$ with some time dependent 
fluctuations. Therefore for the cases where we 
are interested in the long time limit we compute the wavefunction only 
for 70 times values (in the same interval as above with uniform logarithmic 
density) and take the average over the 21 values with 
$10^4\le t/\Delta t\le 10^6$. We note that for the case of a uniform 
wavefunction density the {\em ergodic} values are 
$w_{10,\rm erg}=(21/N)^2$ and $\xi_{\rm IPR,erg}=N^2$. Values of $w_{10}$ 
significantly above $w_{10,\rm erg}$ or of $\xi_{\rm IPR}$ below 
$\xi_{\rm IPR,erg}$ indicate an enhanced probability for the formation of 
compact electron pairs. 

We also mention that both quantities 
$w_{10}$ and $\xi_{\rm IPR}$ are invariant 
with respect to the three 
transformations $p_{+x}\leftrightarrow p_{+y}$, $p_{+x}\to -p_{+x}$ 
and $p_{+y}\to -p_{+y}$ (or $p_{+x}\to 2\pi-p_{+x}$ 
and $p_{+y}\to 2\pi-p_{+y}$) 
corresponding to reflections at the $x$-$y$ diagonal, 
the $y$-axis and the $x$-axis. Even though the effective block Hamiltonian 
(\ref{eq_block_ham}) is not (always) invariant with respect to all three 
of these transformations (see Appendix \ref{appA} for details), the 
choice of an invariant initial state ensures that at finite times 
the wavefunction in block space satisfies for example the identity 
$\bar\psi(p_{+x},p_{+y},\Delta x,\Delta y)=
\bar\psi(p_{+y},p_{+x},\Delta y,\Delta x)$ 
(and similarly for the 
other reflections). In other words a certain reflection transformation 
for $\p_+$ results in the equivalent transformation for the time dependent 
block space wavefunction in $\Dr$ space. Obviously the two quantities 
$w_{10}$ and $\xi_{\rm IPR}$ do not change with respect to these 
transformations (in $\Dr$ space) and therefore they are conserved. 
As a result it is sufficient to compute these quantities only 
for the triangle $0\le p_{+y}\le p_{+x}\le\pi$. 

In the following sections we present the results 
for these quantities and the wavefunction in block representation.

\section{Phase diagram of pair formation}
\label{sec5}

\begin{figure}
  \begin{center}
  \includegraphics[width=0.9\columnwidth]{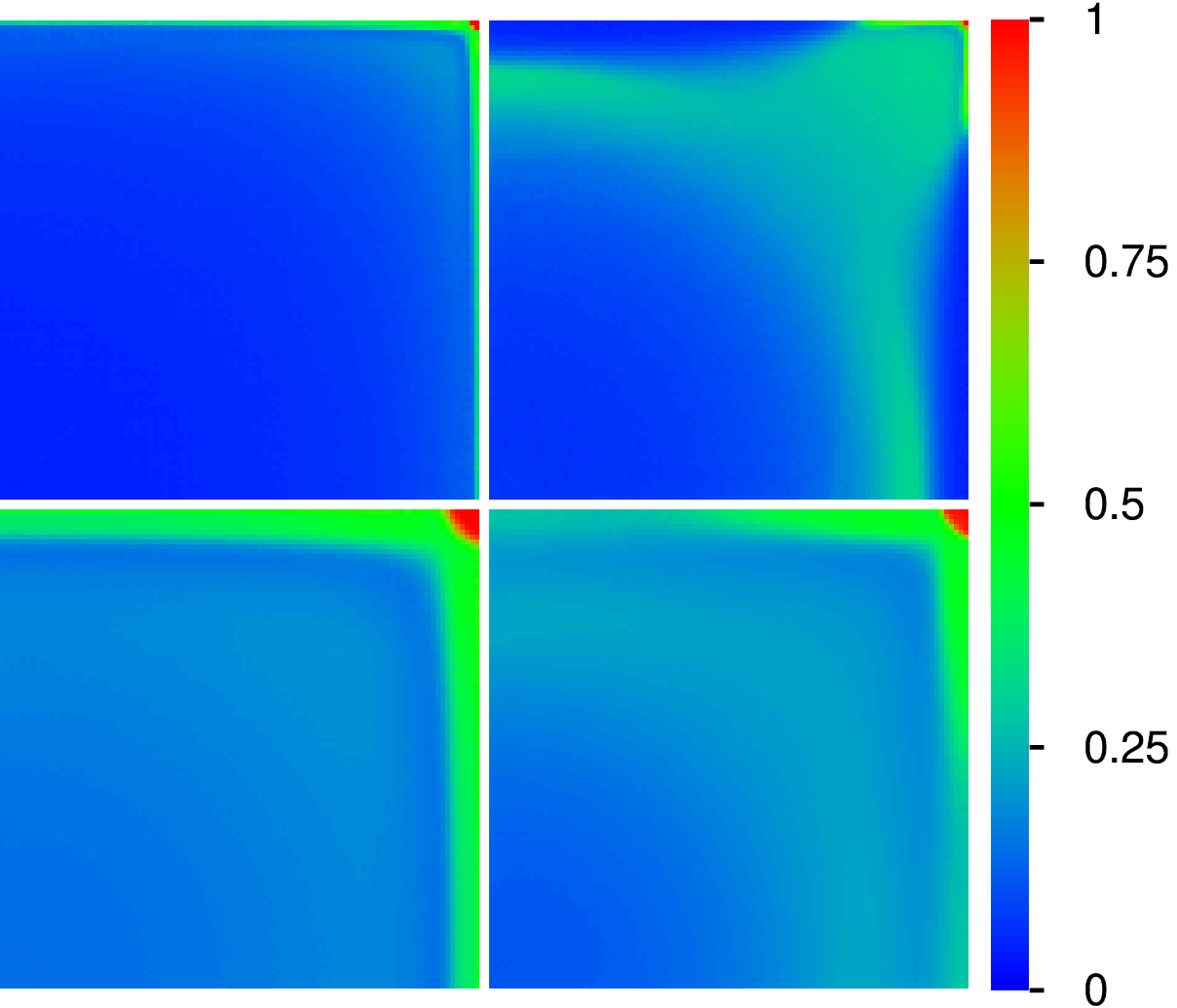}%
  \end{center}
  \caption{\label{fig4}
    Phase diagram of electron pair formation in the plane of pair momentum 
    $\p_+=(p_{+x},p_{+y})$ for the NN-lattice (left panels), the HTC-lattice
    (right panels) and the interaction values $U=0.5$ (top panels), $U=2$ 
    (bottom panels). Shown is the pair formation probability $w_{10}$ 
    for $N=192$ obtained from the exact time evolution for each sector 
    of $\p_+$ with an initial electron distance $\Delta\bar x=\Delta\bar y=1$ 
    and computed from an average over 21 time values in the 
    interval $10^{4}\le t/\Delta t\le 10^{6}$. 
    In all panels the horizontal (vertical) axis corresponds to 
    $p_{+x}\;(p_{+y})\;\in[0,\pi]$ and the numerical values 
    of the color bar correspond to the ratio of $w_{10}$ 
    over its maximal value. The maximum values corresponding to the red 
    region at the top right corner 
    $\p_+=(\pi,\pi)$ are $w_{10}=1$ (both left panels), 
    $w_{10}=0.4510$ (top right) and $w_{10}=0.8542$ (bottom right). 
    For comparison the ergodic value is 
    $w_{10,\rm erg.}=(21/192)^2=0.01196$.
 }
\end{figure}

The phase diagram of the long time average of the pair formation probability 
$w_{10}$ in the $\p_+$-plane is shown in Fig.~\ref{fig4} for both models 
and the interaction values $U=0.5,\,2$. As expected from  the features of the 
effective bandwidth shown in (the top panels of) Fig.~\ref{fig1}, we find 
that globally for both models the pair formation probability is clearly 
maximal at $\p_+=(\pi,\pi)$ and minimal at $\p_+=(0,0)$. Furthermore, 
the size of the maximum region is significantly stronger for $U=2$ than 
for $U=0.5$ which is also to be expected. 
Thus for these $\p_+$ values even a relatively weak or 
moderate Coulomb repulsion creates quite strongly coupled electron pairs.

For the NN-model the top ($p_{+y}=\pi$) or right ($p_{+x}=\pi$) 
boundary also provide large values with $w_{10}\approx 0.5$ 
and the width of these regions is stronger for $U=2$ than 
for $U=0.5$. However, for $U=2$ also the remaining region provides values 
between 0.14 and 0.25 of the maximum value which are clearly above the ergodic 
value 0.012. Even for $U=0.5$ the remaining region is mostly $\approx 0.04$ 
(with some part close to 0.25) which is still above the ergodic 
value.

For the HTC-model the situation is more complicated. The boundary regions 
are more limited, especially for $U=0.5$. However, for the remaining region 
there is a new interesting feature which is 
a significantly enhanced ``green-circle'' of approximate radius
$r_g = \sqrt{ {p^2}_{+x} + {p^2}_{+y}} \approx 0.85 \pi$ for $U=0.5$ 
($w_{10}\approx 0.14$). 
For $U=2$ there is also a circle ($w_{10}\approx 0.20$) with 
approximate radius $r_g\approx 0.75\pi$. 
This circle seems to be less pronounced 
despite its larger value of $w_{10}$ as compared to $U=0.5$ due to the fact 
that the maximum value for $U=2$ ($w_{10}\approx 0.85$ at $\p_+=(\pi,\pi)$) 
is roughly twice the maximum value for $U=0.5$ 
($w_{10}\approx 0.45$). 
This structure cannot be explained by the behavior of the 
effective bandwidth. 
The minimum values of $w_{10}$ at $\p_+\approx (0,0)$ are
$w_{10}\approx 0.02-0.03$ ($w_{10}\approx 0.09-0.10$) for $U=0.5$ ($U=2$) 
which are slightly (significantly) above the ergodic value 0.012.

Globally, nearly for all values of $\p_+$, for both models and both 
interaction values $U=0.5,\,2$ there is an enhanced probability to 
create coupled electron pairs.

\begin{figure}
  \begin{center}
  \includegraphics[width=0.9\columnwidth]{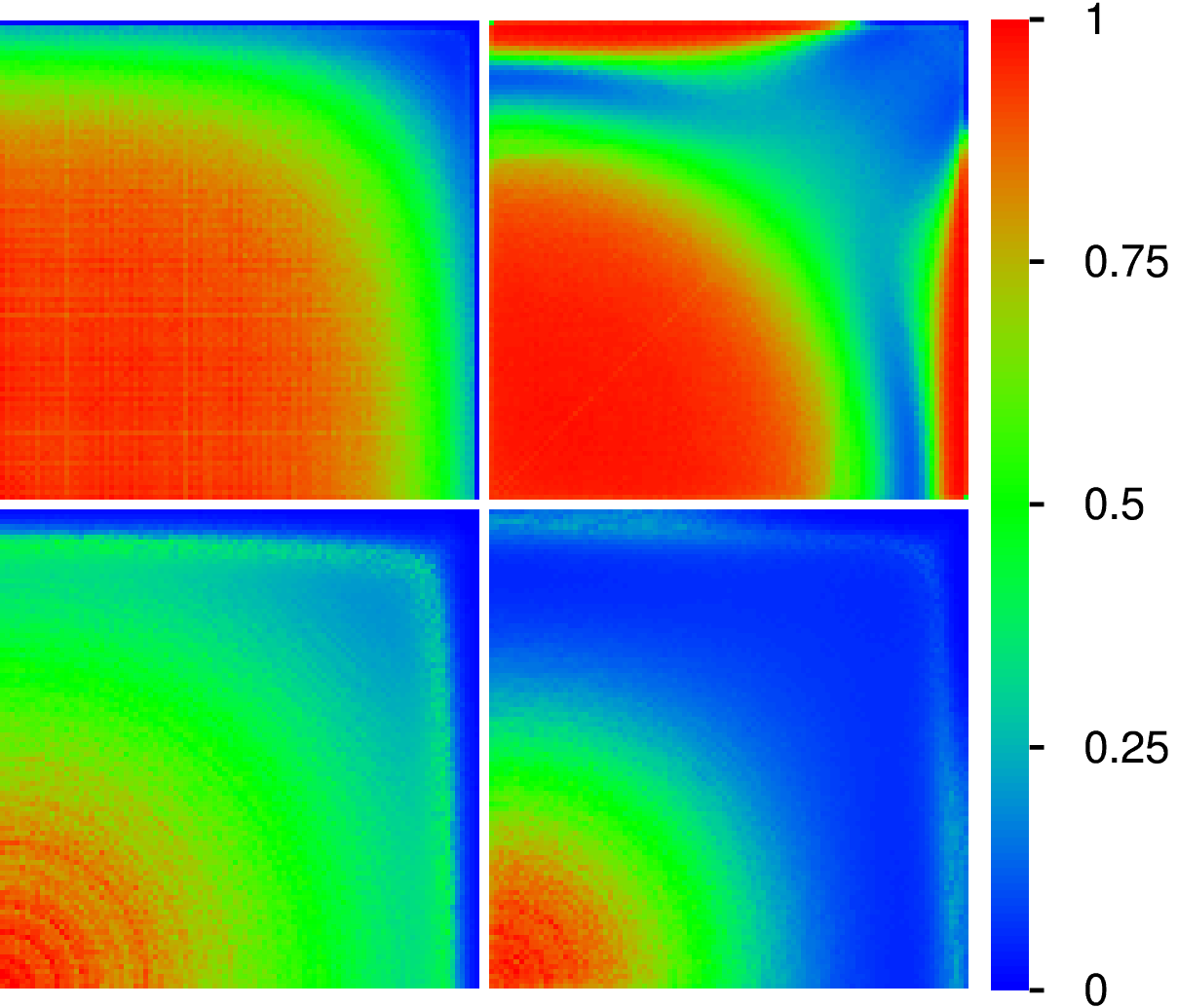}%
  \end{center}
  \caption{\label{fig5}
    Phase diagram of the inverse participation ratio $\xi_{\rm IPR}$ 
    in the plane of pair momentum $\p_+=(p_{+x},p_{+y})$ and computed from 
    the same states, data and cases as in Fig.~\ref{fig4}. 
    The maximum values corresponding to the red 
    region close to the bottom left corner 
    $\p_+=(0,0)$ are $\xi_{\rm IPR}=15300$ (top left), 
    $\xi_{\rm IPR}=4300$ (bottom left), $\xi_{\rm IPR}=18200$ 
    (top right) and $\xi_{\rm IPR}=8600$ (bottom right). 
    The minimum values at the top right corner 
    $\p_+=(\pi,\pi)$ are $\xi_{\rm IPR}=14.87$ (top left), 
    $\xi_{\rm IPR}=4$ (bottom left), $\xi_{\rm IPR}=126$ 
    (top right) and $\xi_{\rm IPR}=9.8$ (bottom right). 
    For comparison the ergodic value is $\xi_{\rm IPR,erg}=192^2=36864$ 
    and the value for the totally symmetrized and localized initial state 
    is $\xi_{\rm IPR,init}=4$. 
  }
\end{figure}

The above observations are perfectly confirmed by the phase diagram for the 
inverse participation ratio $\xi_{\rm IPR}$ which is shown in 
Fig.~\ref{fig5} for the same cases and raw data of Fig.~\ref{fig4}. 
Large (small) values of $\xi_{\rm IPR}$ corresponds to small (large) values 
of $w_{10}$ and a small (strong) pair formation probability. 
Here minimum (maximum) values are at $\p_+=(\pi,\pi)$ ($\p_+=(0,0)$) 
as for the effective bandwidth of Fig.~\ref{fig1} (see figure caption 
for the numerical minimum, maximum and ergodic values). The boundary 
structure of the NN-model and the circle-structure of the HTC-case are 
also clearly visible. 

\begin{figure}
  \begin{center}
  \includegraphics[width=0.9\columnwidth]{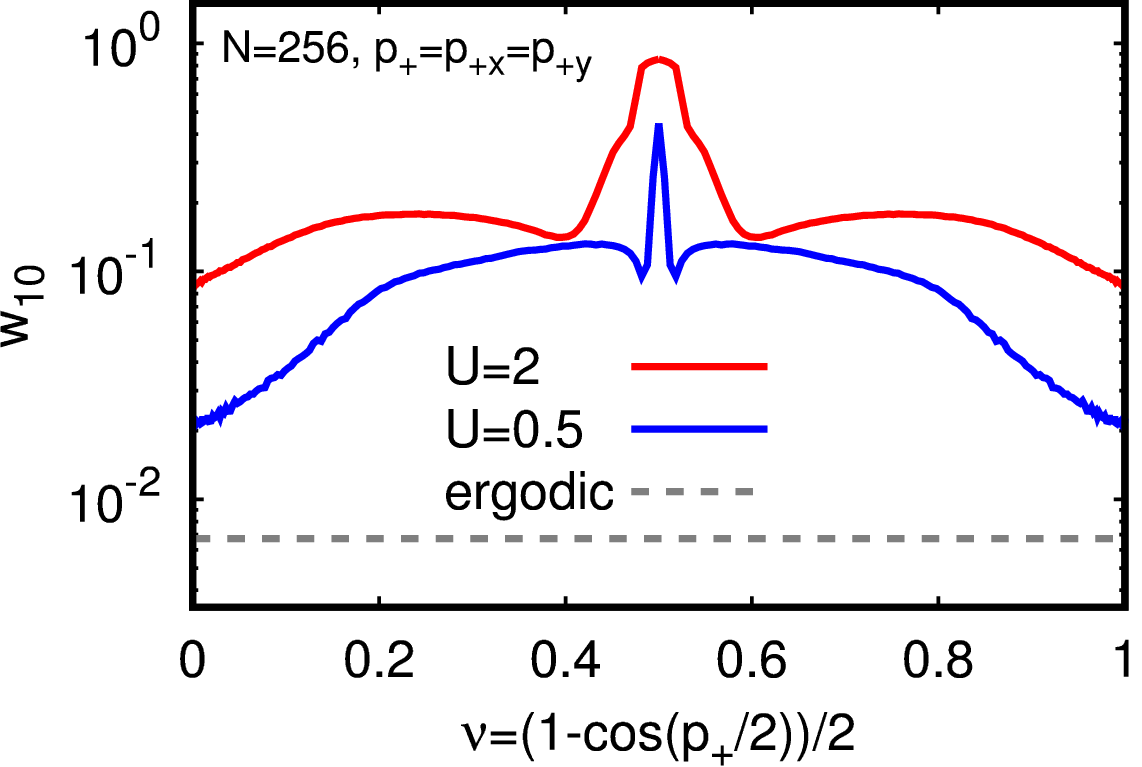}%
  \end{center}
  \caption{\label{fig6}
    Dependence of the electron pair formation probability $w_{10}$ on 
    $\nu=(1-\cos(p_+/2))/2$ for $p_+= p_{+x}=p_{+y}$ and the HTC-model 
    at $U=0.5,\,2$ and $N=256$. 
    $w_{10}$ is computed from the same long time average as in 
    Fig.~\ref{fig4}. The maximum value at 
    $\nu=\nu_{\rm max}=0.5$ is $w_{10}=0.8535$ ($w_{10}=0.4456$) for 
    $U=2$ ($U=0.5$).
    See Fig. 5 of \cite{prr2020} for the corresponding figure for the 
    NN-model. For the NN-model the maximum value at $\nu=\nu_{\rm max}=0.5$ 
    is exactly $w_{10}=1$ for both interaction values.
  }
\end{figure}

We have also computed the long time average of the pair formation probability 
for the HTC-model at larger system size $N=256$ and the special cases of 
either $p_{+x}=p_{+y}$ or $p_{+y}=0$ where the additional second symmetry (see 
discussion in the previous section and Appendix \ref{appA}) 
reduces the computational effort. In this way we can explore the diagonal 
and right boundary of the phase diagram in more detail. 

Figure~\ref{fig6} shows $w_{10}$ for the HTC-model, $N=256$, 
$p_+=p_{+x}=p_{+y}$ 
and both interaction values $U=0.5,\,2$ as a function of the 
parameter $\nu=(1-\cos(p_+/2))/2$. Both curves clearly confirm some of the 
observations of the phase diagrams, i.e. strongest pair formation probability 
at $\nu=0.5$ ($p_{+x,y}=\pi$) with a somewhat larger 
maximum range for $U=2$ as compared to $U=0.5$ and 
a minimal pair formation probability at $\nu=0$ ($p_{+x,y}=0$) 
or $\nu=1$ ($p_{+x,y}=2\pi$) but still clearly above the ergodic limit for 
all cases. 
The precise numerical maximum values of $w_{10}$ at $\nu=0.5$ are slightly 
different from, but still in general agreement with, those 
of Fig.~\ref{fig4} due to the different system size.
The corresponding figure for the NN-model was already given in \cite{prr2020}. 

\begin{figure}
  \begin{center}
  \includegraphics[width=0.9\columnwidth]{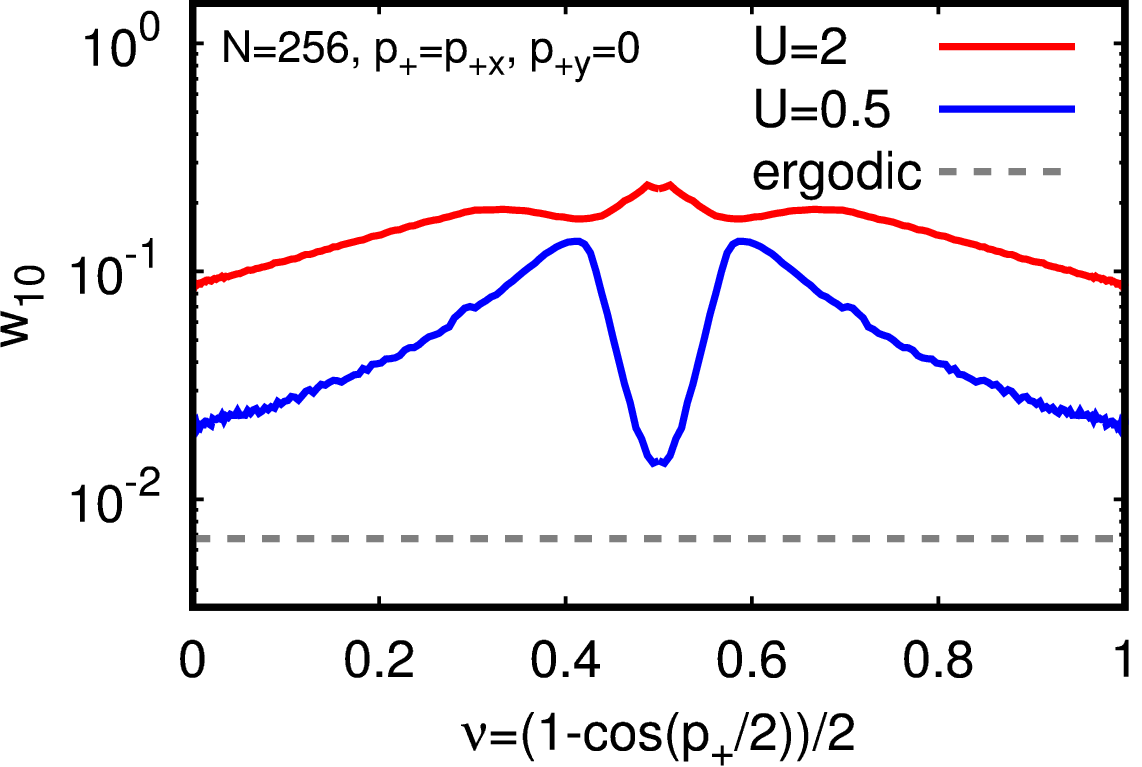}%
  \end{center}
  \caption{\label{fig7}
    Dependence of the electron pair formation probability $w_{10}$ on 
    $\nu=(1-\cos(p_+/2))/2$ for $p_+= p_{+x}, p_{+y}=0$ and the HTC-model
    at $U=0.5, 2$ and $N=256$. $w_{10}$ is computed from the same 
    long time average as in Fig.~\ref{fig4}. 
    The value at $\nu=0.5$ is $w_{10}=0.2302$ ($w_{10}=0.01479$) for 
    $U=2$ ($U=0.5$).
  }
\end{figure}

Figure~\ref{fig7} shows $w_{10}$ for the HTC model, $N=256$ and 
both interaction values $U=0.5,\,2$ at the boundary $p_{+y}=0$ 
as a function of the parameter $\nu=(1-\cos(p_+/2))/2$ with $p_+=p_{+x}$. 
The curve for $U=0.5$ clearly shows a strong local maximum at 
$\nu\approx 0.5\pm 0.1$ ($p_+\approx \pi\pm \pi/8$) corresponding to 
green-circle with radius $r_g\approx 0.85\pi$ visible in the phase diagram. 
For $U=2$ there are higher but less pronounced local maxima at 
$\nu\approx 0.5\pm 0.19$ corresponding to the slightly visible circle for this 
case. However, at $U=2$ the value of $w_{10}$ at $\nu=0.5$ is rather high 
while at $U=0.5$ its value at $\nu=0.5$ is quite low but still clearly 
above the ergodic limit. 

Figures~S1 and S2 of the supplementary material are similar to 
Figs.~\ref{fig6} and \ref{fig7} respectively but for the inverse 
participation ratio $\xi_{\rm IPR}$. 

\section{Time evolution of pair formation}
\label{sec6}

\begin{figure}
  \begin{center}
  \includegraphics[width=0.9\columnwidth]{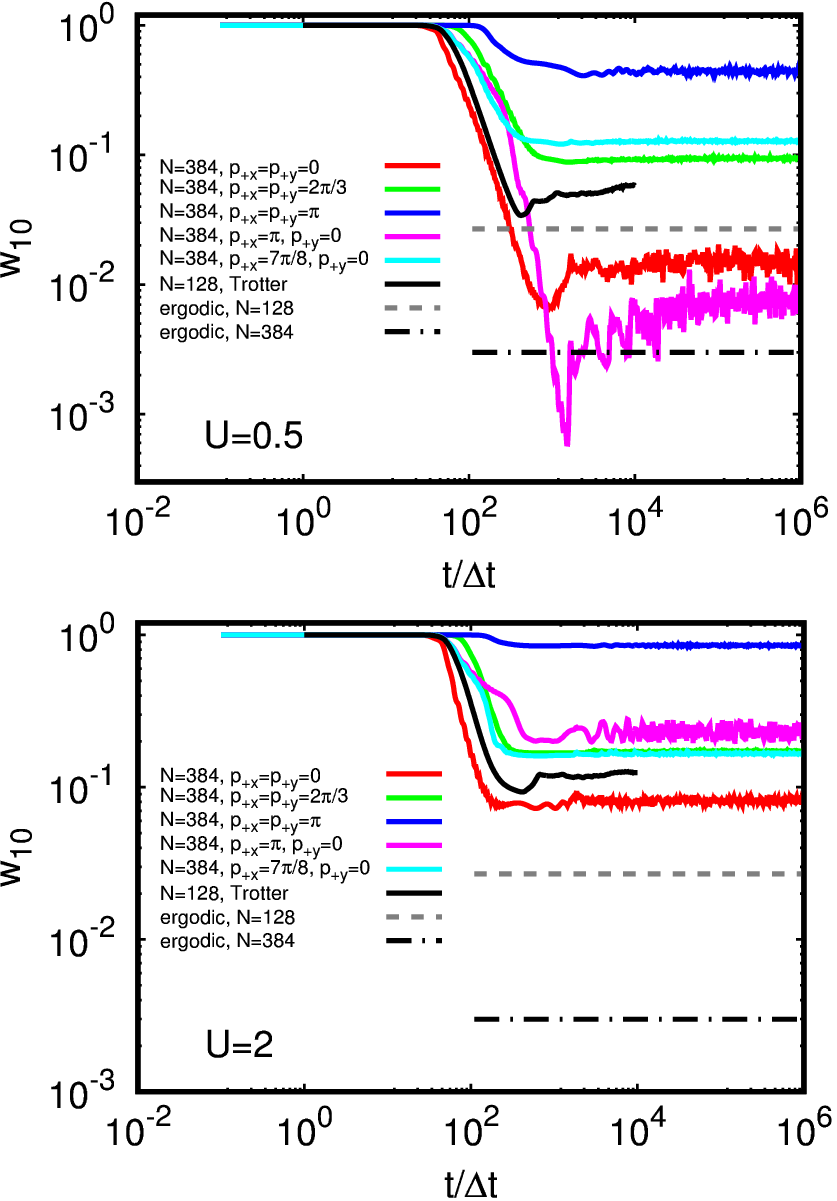}%
  \end{center}
  \caption{\label{fig8}
    Time dependence of the pair formation probability $w_{10}$ 
    for $U=0.5$ (top panel) and $U=2$ (bottom panel) and different 
    cases of the exact time evolution in certain 
    $\p_+=(p_{+x},p_{+y})$ sectors at 
    $N=384$ and the full space Trotter formula time evolution at $N=128$. 
    The dashed lines correspond to the ergodic values $(21/N)^2=0.0269$ for 
    $N=128$ (grey dashed) and $(21/N)^2=0.00299$ for $N=384$ (black dashed).
  }
\end{figure}

We also computed a more precise time evolution of the pair formation 
probability $w_{10}$ for the larger system size $N=384$ and certain 
specific cases $p_{+x}=p_{+y}\;\in\;\{0,\,2\pi/3,\,\pi\}$ and 
$p_{+y}=0$ with $p_{+x}\;\in\;\{7\pi/8,\,\pi\}$. The results together 
with the full space results using the Trotter formula approximation 
at $N=128$ are shown in Fig.~\ref{fig8} for $U=0.5,\,2$. 
In all cases the value of $w_{10}$ starts decaying from its initial 
value $w_{10}=1$ at $t/\Delta t>20$-$30$ and converges to a long time 
saturation value for $t/\Delta t>10^3$ sometimes with some temporal 
quasi-periodic fluctuations. 
In most cases the saturation values at $U=2$ are clearly larger than for 
$U=0.5$ except for the case $p_{+y}=0$ and $p_{+x}=7\pi/8$ where both 
saturation values are somewhat comparable. In particular, at $U=0.5$ 
the value for $p_{+y}=0$ and $p_{+x}=7\pi/8$ is significantly larger 
than the value for $p_{+y}=0$ and $p_{+x}=\pi$ while at $U=2$ 
it is the inverse. 
This observation is in agreement with the appearance of the green circle in 
the phase diagram where for $U=0.5$ the circle is dominant in comparison 
to the right boundary while for $U=2$ it is dominated by the right boundary. 

The saturation value of the data obtained by the 
Trotter formula approximation, 
which somehow corresponds to an average over all possible $\p+$ values,  
is quite low if compared to the case $\p_+=0$ but still clearly above the 
corresponding ergodic value (for its reduced system size). Also for most of 
the other cases the saturation value is clearly above the ergodic value except 
for $U=0.5$, $p_{+y}=0$ and $p_{+x}=\pi$ where the curve is a 
$t\approx 10^3\Delta t$ even below the ergodic value and saturates later 
at a value only slightly above the ergodic value. 

\begin{figure}
  \begin{center}
  \includegraphics[width=0.9\columnwidth]{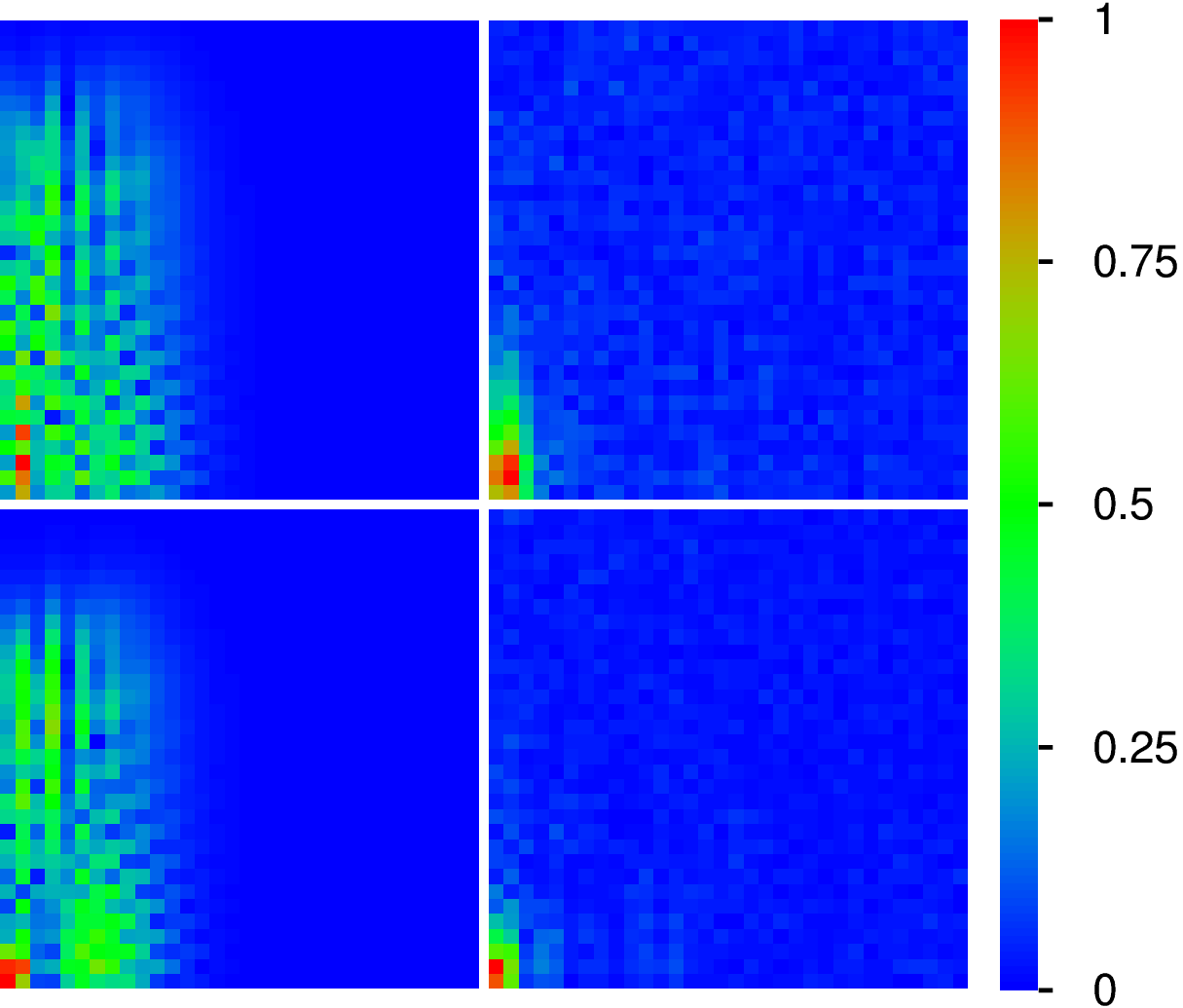}%
  \end{center}
  \caption{\label{fig9}
    Color plot of wavefunction amplitude 
    $|\bar\psi(\p_+,\Dr)|$ in block representation in 
    $\Dr=(\Delta x,\Delta y)$ plane 
    obtained from the 2D quantum time evolution for the HTC lattice 
    with $N=384$ and the sector $p_{+x}=7\pi/8$, $p_{+y}=0$.
    All panels show a zoomed region $0\le \Delta x,\Delta y<32$.
    Left (right) panels correspond to 
    $t=100\,\Delta t$ ($t=10^5 \Delta t$; with $\Delta t=1/B_2=1/(16+U)$) 
    and top (bottom) panels 
    correspond to interaction strength $U=0.5$ ($U=2$). 
    Related videos are available at \cite{ourwebpage}.
 }
\end{figure}

Motivated by the observation of the green-circle at radius 
$r_g\approx 0.85\pi$ in the phase diagram for $U=0.5$, we show in 
Fig.~\ref{fig9} the wavefunction amplitude at $p_{+x}=7\pi/8$, $p_{+y}=0$, 
$N=384$ and both interaction values $U=0.5,\,2$ and two time values 
$t/\Delta t=100,\,10^5$. The first observation is that the diffusive 
spreading in $x$-direction is strongly suppressed if compared to the 
$y$-direction which is expected since $p_{+x}$ is rather close to $\pi$ 
while $p_{+y}=0$. 

At $U=0.5$ the steady-state  at $t/\Delta t=10^5$, despite a smaller 
value of $w_{10}=0.0754$ if compared to $w_{10}=0.1342$ at $U=2$, 
has a larger spatial extension of $\sim 30$ lattice sites compared 
to $\sim 12$ lattice sites for $U=2$. This in rough qualitative 
agreement with the values $\xi_{\rm IPR}=940$ (for $U=0.5$) and 
$\xi_{\rm IPR}=268$ (for $U=2$). However, a large amount of the contribution 
to the inverse participation ratio comes from the remaining probability of 
about 87-90\% which has uniformly spread over the full lattice thus explaining 
the difference between $\xi_{\rm IPR}$ and the visible spatial extension 
in Fig.~\ref{fig9} (for this reason with consider $w_{10}$ to be a more 
suitable quantity than $\xi_{\rm IPR}$ to describe the pair formation 
probability).

\section{Results overview}
\label{sec7}

\begin{figure}
  \begin{center}
  \includegraphics[width=0.9\columnwidth]{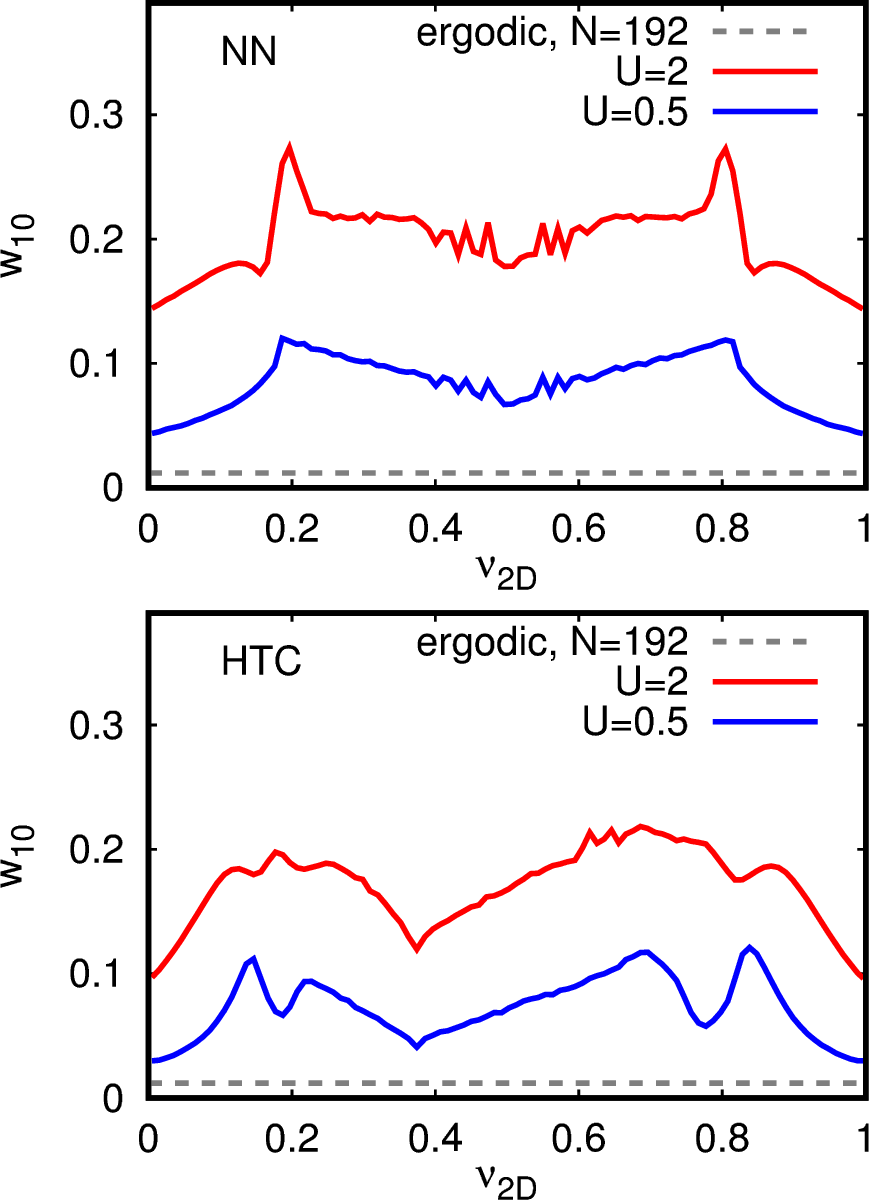}%
  \end{center}
  \caption{\label{fig10}
    Dependence of the electron pair formation probability $w_{10}$ on 
    the effective 2D filling factor $\nu_{2D}$ for the NN-lattice (top) 
    and the HTC-lattice (bottom). The values of $w_{10}$ have been obtained 
    from the data of Fig.~\ref{fig4} (for $N=192$) by an average along lines 
    of constant electron pair energy $E_c$ at momenta $\p_1=\p_2=\p_+/2$ 
    with $p_{+x},\,p_{+y}\in[0,2\pi]$. Lowest (largest) energy corresponds 
    to $\nu_{2D}=0$ ($\nu_{2D}=1$). The data points shown correspond to an 
    effective histogram with bin width $\Delta \nu_{2D}\approx 0.01$. 
    The red (blue) curve corresponds to the interaction value $U=2$ ($U=0.5$) 
    and the grey dashed line corresponds to the ergodic value 
    $(21/192)^2=0.01196$.
 }
\end{figure}

The discussion of the phase diagram given in Fig.~\ref{fig4} has shown that 
the pair formation probability is maximal at the point $\p_+=(\pi,\pi)$. 
However, the surrounding region to this point is quite small if compared 
to the green-circle where we have a somewhat more modest pair 
formation probability. In terms of available values of $\p_+$ the latter 
region is possibly more important. In order to analyze this point in a more 
quantitative way, we assume a simple model where both electrons have 
the same momentum $\p_+/2$ (i.e. $\Delta p=0$) and where the available states 
of this type are filled from smallest to largest energies. We subdivide 
these states, ordered in energy, in slices of equal number ($\sim 1/100$ of 
all available states) and compute the average of $w_{10}$ for each slice 
which is equivalent to the average of $w_{10}$ at lines of constant energy. 
In Fig.\ref{fig10}, we show the dependence of this average on the 
effective 2D-filling factor $\nu_{2D}$ which is the weight of slices 
below a certain energy. 

For the NN-model we observe a strong peak at $\nu_{2D}=0.2$ 
(and similarly at $\nu_{2D}=0.8$ due to symmetry). This peak is caused 
by the combination of the maximum point at $\p_+=(\pi,\pi)$ and 
rather strong (top or right) boundary contributions visible in the 
left panels of Fig.~\ref{fig4}. For the HTC-model at $U=2$ this peak is 
still visible but its value is reduced. However, for $U=0.5$, there are 
two separated peaks, a stronger one at $\nu_{2D}\approx 0.15$ related 
to the average over the green circle at radius $r_g\approx 0.85$ and 
a second lower peak at $\nu_{2D}\approx 0.24$ related to the average 
of the maximum region close to $\p_+=(\pi,\pi)$. For this particular case, 
the green circle has a stronger global contribution to the pair formation 
probability than the maximum region at $\p_+=(\pi,\pi)$. 

\section{Discussion}
\label{sec8}

In our studies we analyzed the electron pair formation 
in a tight-binding model of
La-based cuprate superconductors induced by Coulomb repulsion.
Our analytical and numerical results show that even a 
repulsive Coulomb interaction 
can form two electron pairs with a high probability.
Such pairs have a compact size and propagate through the whole system.
We expect that such pairs may contribute to the emergence of 
superconductivity in La-based cuprates.

Of course, our analysis only considers two electrons and 
in a real system at finite electron density there is a Fermi sea
which can modify electron interactions. However, we expect 
that electrons significantly below the Fermi energy 
will only create a mean-field potential which will
not significantly affect interacting electrons
with energies in the vicinity of the Fermi energy.
A detailed investigation of effects of finite electron density
on the Coulomb pair formation represents an important
task for future studies.

\section{Acknowledgments}
This work has been partially supported through the grant
NANOX $N^o$ ANR-17-EURE-0009 in the framework of 
the Programme Investissements d'Avenir (project MTDINA).
This work was granted access to the HPC resources of 
CALMIP (Toulouse) under the allocation 2020-P0110.

\appendix
\section{Appendix}
\label{appA}

In this appendix we present the derivation of the block Hamiltonian 
(\ref{eq_block_ham}) and a more detailed discussion about its discrete 
symmetries. In order to simplify the notations, we will use here the 
full set $\A'=\A\cup(-\A)$ of neighbor vectors (in the full and not only 
half plane) for the summation over the vectors $\a$ which allows to reduce 
the number of terms in the following expressions. The TIP Hamiltonian 
(\ref{eq_quant_Ham}) can then be written in a more explicit form as:
\begin{eqnarray}
\nonumber
H&=&-\sum_{\r_1,\r_2}\sum_{\a\in\A'} 
t_\a\bigl(\ket{\r_1,\r_2}\bra{\r_1+\a,\r_2}
+\ket{\r_1,\r_2}\bra{\r_1,\r_2-\a}\bigr)\\
\label{eq_H_explicit}
&&+\sum_{\r_1,\r_2}\bar U(\r_2-\r_1)\ket{\r_1,\r_2}\bra{\r_1,\r_2}
\end{eqnarray}
where for convenience we have written ``$\r_2-\a$'' instead 
of ``$\r_2+\a$'' (in the second term 
of the first line) since for $\a\in\A'$ also 
$-\a\in\A'$. Furthermore, the terms with shifts of $\a$ in the left side have 
been absorbed by the increased set $\A'$ (with respect to $\A$ used in 
(\ref{eq_H1p})) combined with a subsequent shift of the 
summation index $\r_1$ or $\r_2$ and exploiting the periodic boundary 
conditions. 

Applying (\ref{eq_H_explicit}) to a block basis state (\ref{eq_block_basis}) 
we find that:
\begin{eqnarray}
\nonumber
H\ket{\p_+,\Dr}&=&-\frac{1}{N}\sum_{\r_1}\sum_{\a\in\A'} 
t_\a\Bigl(\ket{\r_1-\a,\r_1+\Dr}\,e^{i\p_+\cdot(\r_1+\Dr/2)}\\
\label{eq_block1_applied}
&&+\ket{\r_1,\r_1+\Dr+\a}\,e^{i\p_+\cdot(\r_1+\Dr/2)}\Bigr)\\
\nonumber
&&+\bar U(\Dr)\ket{\p_+,\Dr}.
\end{eqnarray}

Using the shift $\r_1\to \r_1+\a$ in the $\r_1$-sum of the first line of this 
expression we obtain:
\begin{eqnarray}
\nonumber
H\ket{\p_+,\Dr}&=&-\frac{1}{N}\sum_{\r_1}\sum_{\a\in\A'} 
t_\a\Bigl(\ket{\r_1,\r_1+\Dr+\a}\,e^{i\p_+\cdot(\r_1+\a+\Dr/2)}\\
\label{eq_block2_applied}
&&+\ket{\r_1,\r_1+\Dr+\a}\,e^{i\p_+\cdot(\r_1+\Dr/2)}\Bigr)\\
\nonumber
&&+\bar U(\Dr)\ket{\p_+,\Dr}
\end{eqnarray}
which can be rewritten as:
\begin{eqnarray}
\nonumber
H\ket{\p_+,\Dr}&=&-\frac{1}{N}\sum_{\r_1}\sum_{\a\in\A'} 
t_\a\,\ket{\r_1,\r_1+\Dr+\a}\\
\nonumber
&&\qquad \times\,e^{i\p_+\cdot[\r_1+(\Dr+\a)/2]}
\underbrace{\left(e^{i\p_+\cdot\a/2}+e^{-i\p_+\cdot\a/2}\right)}_
{2\cos(\p_+\cdot \a/2)}\\
\label{eq_block3_applied}
&&+\bar U(\Dr)\ket{\p_+,\Dr}\\
\label{eq_block4_applied}
&=&-2\sum_{\a\in\A'}t_\a\,\cos(\p_+\cdot \a/2)\,
\ket{\p_+,\Dr+\a}\\
\nonumber
&&+\bar U(\Dr)\ket{\p_+,\Dr}\ .
\end{eqnarray}
The last expression provides exactly the effective block Hamiltonian 
(\ref{eq_block_ham}) if we replace the sum over 
$\a\in\A'$ by a sum over $\a\in\A$ with two contributions ``$+\a$'' 
and ``$-\a$'' and applying for the latter contribution 
a subsequent shift $\Dr\to\Dr+\a$ in the $\Dr$ sum. 
However, there is one additional complication if 
$\Dr+\a=(\Delta x+a_x,\Delta y+a_y)$ in (\ref{eq_block4_applied}) 
leaves the initial square of 
$\Delta x,\Delta y\in\{0,\ldots N-1\}$. Then we have to add (subtract)
$N$ to (from) $\Delta x+a_x$ and/or $\Delta y+a_y$ which provides according to 
(\ref{eq_block_basis}) the factor 
$e^{\pm ip_{+x}N/2}=e^{\pm i\pi l_{+x}}=(-1)^{l_{+x}}$ (for $\Delta x$ and 
similarly for $\Delta y$) resulting in either periodic or anti-periodic 
boundary conditions in $x$- ($y$-)di\-rection 
depending on the parity of the integer index $l_{+x}$ ($l_{+y}$). 

We close this appendix with a short discussion about the discrete 
reflection symmetries of the block Hamiltonian (\ref{eq_block_ham}) and 
the possibility to reduce its effective matrix size $N^2$ due to such 
symmetries. For the NN-model, as already 
discussed in detail in \cite{prr2020}, there are at least two symmetries 
with respect to $\Delta x\to N-\Delta x$ (reflection at the 
$\Delta y$-axis) or $\Delta y\to N-\Delta y$ (reflection at the 
$\Delta x$-axis) and 
in case if $p_{+x}=p_{+y}$ there is a third symmetry with respect to 
$\Delta x\leftrightarrow\Delta y$ 
(reflection at the $\Delta x$-$\Delta y$ diagonal) 
which allows for an effective matrix size of roughly 
either $N^2/4$ or $N^2/8$ (if $p_{+x}=p_{+y}$). 

However, for a more general 
lattice, such as the HTC-model, or more generally in presence of at least one 
neighbor vector $\a=(a_x,a_y)$ with both $a_x\neq 0$ and $a_y\neq 0$ (e.g. 
$\a=(1,1)$) the number of symmetries is reduced. 
For the most generic case with $p_{+x}\neq p_{+y}$, $p_{+x}\neq 0$ and 
$p_{+y}\neq 0$ there is only one symmetry corresponding to particle exchange 
with two simultaneous transformations $\Delta x\to N-\Delta x$ {\bf and} 
$\Delta y\to N-\Delta y$ which allows for 
a reduction of the effective matrix size 
to $\approx N^2/2$. In this case the factors 
$\cos(\p_+\cdot \a/2)=\cos[(p_{+x}a_x+p_{+y}a_y)/2]$ appearing in the 
effective hopping amplitudes are not modified because the replacement 
$\a\to -\a$ due the symmetry transformation only changes the global 
sign inside the cosine argument. However, this is no longer true 
if we apply for example the transformation $\Delta x\to N-\Delta x$ 
without modifying $\Delta y$ which is equivalent to the 
replacement of $(a_x,a_y)\to(-a_x,a_y)$ of the neighbor vectors. Therefore 
a {\bf single} reflection at the $\Delta y$ (or $\Delta x$) axis modifies 
the hopping amplitude (if both $a_x\neq 0$, $a_y\neq 0$ and also both 
$p_{+x}\neq 0$, $p_{+y}\neq 0$) 
and (\ref{eq_block_ham}) is (in general) 
not invariant with respect to such transformations. However, if either 
$p_{+x}=0$ or $p_{+y}=0$ the effective hopping amplitudes are not modified 
with respect to these two individual reflections and we have two symmetries 
with an effective matrix size of $\approx N^2/4$. 
Also if $p_{+x}=p_{+y}\neq 0$ we have two symmetries (particle exchange 
and reflection at the $\Delta x$-$\Delta y$ diagonal) leading also to an 
effective matrix size of $\approx N^2/4$. Finally, for the special 
case $p_{+x}=p_{+y}=0$, we have even three symmetries (as in the NN-Model 
for $p_{+x}=p_{+y}$) with effective matrix size of $\approx N^2/8$.

\setcounter{figure}{0} \renewcommand{\thefigure}{S\arabic{figure}} 
\setcounter{equation}{0} \renewcommand{\theequation}{S\arabic{equation}} 
\setcounter{page}{1}

\noindent{{\bf Supplementary Material for\\
\vskip 0.2cm
\noindent{Coulomb electron pairing in a tight-binding model of La-based cuprate superconductors}}\\
\noindent by
K.~M.~Frahm and D.~L.~Shepelyansky.}

\vspace{0.5cm}
Here, we present additional material for the main part of the article.

\begin{figure}
  \begin{center}
  \includegraphics[width=0.9\columnwidth]{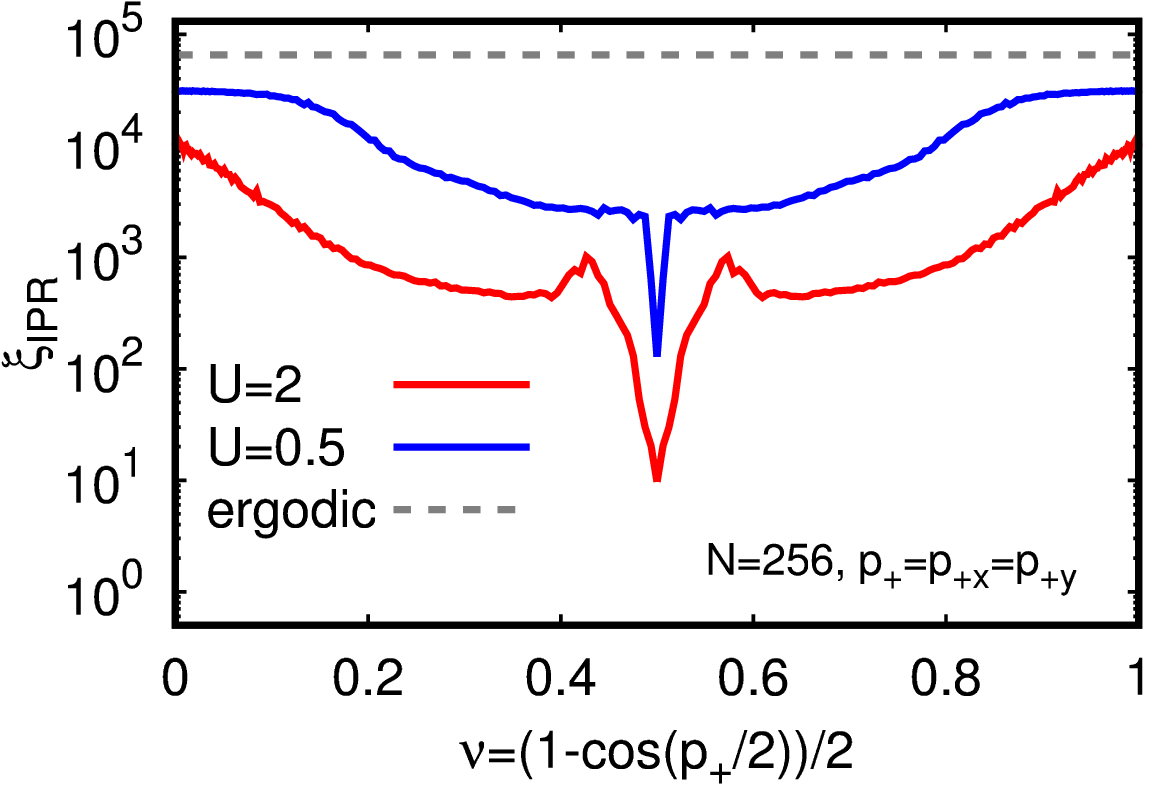}%
  \end{center}
  \caption{\label{figS1}
    As Fig.~6 but for the inverse participation ratio
    $\xi_{\rm IPR}$. 
  }
\end{figure}

Figure~\ref{figS1} presents data for the inverse participation ratio
for the case of Fig.~6.

\begin{figure}
  \begin{center}
  \includegraphics[width=0.9\columnwidth]{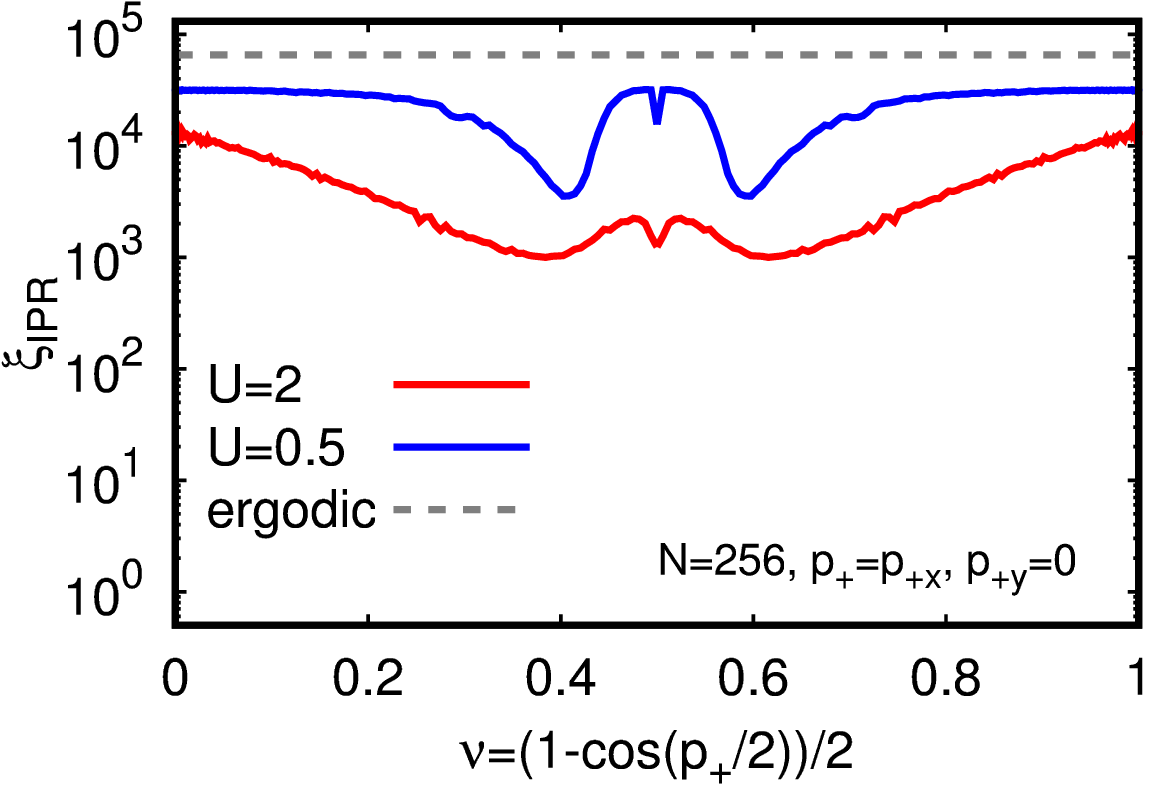}%
  \end{center}
  \caption{\label{figS2}
    As Fig.~7 but for the inverse participation ratio
    $\xi_{\rm IPR}$. 
  }
\end{figure}

Figure~\ref{figS2} presents data for the inverse participation ratio
for the case of Fig.~7.

Two video files for the time evolution obtained by the 
Trotter formula approximation corresponding to the parameters 
of Fig.~2 and Fig.~3
are presented in files \texttt{videofig2.avi}
for the density $\rho_{XX}(x_1,x_2)$ 
defined in Eq.~(8) and in \texttt{videofig3.avi}
for the density $\rho_{\rm rel}(\Delta x,\Delta y)$ 
defined in Eq.~(9) (here $N=128$, $U=2$). 
Both video files provide a direct comparison 
between the NN-model (right box in video) and the HTC-model (left box in 
video) and correspond to 464 time values $t=l_j\,\Delta t$ (25 values per 
second of video) with 
integer $l_0=0$, $1\le l_j\le 10^4$ for $j=1,\ldots, 463$ and roughly 
uniform logarithmic density.


\end{document}